\documentclass[10pt,aps,prl,twocolumn,superscriptaddress]{revtex4}

\usepackage{graphicx}
\usepackage{amsmath,amssymb}
\usepackage{color}
\usepackage{url}
\usepackage{xfrac}
\usepackage{upgreek}
\usepackage[outdir=./]{epstopdf}
\usepackage{siunitx}

\graphicspath{{./Images/}}

\usepackage[linktocpage=true, colorlinks=true, urlcolor=blue, linkcolor=blue, citecolor=blue]{hyperref}
\usepackage{cleveref}

\begin{document}

\title{Measuring choriocapillaris blood flow with laser Doppler optical coherence tomography}

\author{L\'eo Puyo}
\affiliation{Institute of Biomedical Optics, University of L\"ubeck, Peter-Monnik-Weg 4, 23562 L\"ubeck, Germany}
\affiliation{gl.puyo@gmail.com}

\author{Jonas Franke}
\affiliation{Institute of Biomedical Optics, University of L\"ubeck, Peter-Monnik-Weg 4, 23562 L\"ubeck, Germany}

\author{Lisa Kutzner}
\affiliation{Heidelberg Engineering GmbH, Rapsacker 1, 23556 L\"ubeck, Germany}

\author{Clara Pf\"affle}
\affiliation{Institute of Biomedical Optics, University of L\"ubeck, Peter-Monnik-Weg 4, 23562 L\"ubeck, Germany}

\author{Hendrik Spahr}
\affiliation{Institute of Biomedical Optics, University of L\"ubeck, Peter-Monnik-Weg 4, 23562 L\"ubeck, Germany}

\author{Gereon H\"uttmann}
\affiliation{Institute of Biomedical Optics, University of L\"ubeck, Peter-Monnik-Weg 4, 23562 L\"ubeck, Germany}
\affiliation{Medical Laser Center L\"ubeck GmbH, Peter-Monnik-Weg 4, 23562 L\"ubeck, Germany}
\affiliation{Airway Research Center North (ARCN), Member of the German Center for Lung Research (DZL)}


\begin{abstract}
We report on using a laser Doppler processing of Fourier-domain optical coherence tomography (OCT) data for the assessment of pulsatile blood flow in the choriocapillaris. Signal fluctuations in B-scans recorded at 2 kHz were analyzed by Fourier transform to extract blood flow information. The spectral broadening of light backscattered by the choriocapillaris was used to derive a choriocapillaris flow velocity index in physical units, with sufficient temporal resolution to capture heartbeat-induced variations. Furthermore, the asymmetry in the spectral broadening enabled us to determine the axial direction of blood flow with high sensitivity, allowing for the detection of flow orientation in retinal capillaries. This approach is promising as it can be directly implemented on widely available fast-scanning Fourier-domain OCT instruments.
\end{abstract}

\maketitle

The choriocapillaris supplies photoreceptors and the retinal pigment epithelial (RPE) cells. It is of high interest for many prevalent ocular diseases including age-related macular degeneration, diabetic retinopathy, myopia, and glaucoma~\cite{Lejoyeux2022}. Due to its location below the RPE, the choriocapillaris is challenging to image \textit{in vivo}. Progress in optical coherence tomography (OCT) in the last decade has however allowed significant improvements.
OCT techniques for blood flow measurement generally fall into two categories, based on different physical principles: Doppler shift and speckle decorrelation~\cite{Srinivasan2012d}.

Measuring the Doppler phase shift caused by axial movement enables measuring the exact flow when knowing the vessels’ diameter and inclination to the optical axis~\cite{Leitgeb2014}. Doppler OCT can therefore detect the axial direction of blood flow~\cite{Izatt1997}. The detection of only two time points is necessary to measure the Doppler phase shift, but adverse effects happen when the velocity reaches the maximal measurable velocity due to phase wrapping. Phase-resolved Doppler OCT faces the challenge that the angle detection is critical and small angular changes occur with eye motion and lead to imprecision in flow measurements. Despite the development of approaches such as multi-beam imaging and adaptive scanning patterns to address these challenges~\cite{Haindl2016, Desissaire2020}, the added technical complexity has hindered clinical adoption. Finally, Doppler OCT is generally considered unsuited to measure capillary blood flow because phase shifts are small and fall into the phase-noise, making Doppler OCT hardly suitable to study the choriocapillaris.

Capillary blood flow can be efficiently imaged through speckle decorrelation, which is caused by the displacement of scatterers relatively to the imaging voxel, and leads to a spectral broadening. Unlike Doppler-OCT, this approach can detect flow in vessels perpendicular to the optical axis. Blood flow contrasts can be obtained through speckle variance, and metrics such as decorrelation time and spectrum width can be used to quantify absolute flow in physical units~\cite{Srinivasan2010}. These estimations are however less direct than those obtained from Doppler shifts. A popular OCT approach relying on speckle decorrelation is OCT-angiography (OCT-A), where calculating the standard deviation of a few B-scans provides high resolution angiographies of the retina~\cite{Makita2006, Wang2007}. OCT-A images can reveal the choriocapillaris meshwork~\cite{Migacz2019}, but only qualitatively, as they are limited in temporal frequency resolution due to insufficient sampling. While OCT-A can provide information on the morphology of the choriocapillaris using metrics such as vascular density and flow voids~\cite{Zhang2018}, these metrics describe the vascular structure rather than the actual perfusion of the vessels.


In this study, we employed a high-speed Fourier-domain scanning swept-source OCT device to characterize the choriocapillaris perfusion and study its pulsatile behavior. To circumvent the limitation of OCT-A, we captured B-scans continuously at a fixed location, enabling the sampling of both low- and high-frequencies of the spectrum broadening. Algorithms developed for laser Doppler holography (LDH) were applied to measure dynamic changes in the spectrum of light scattered in the choriocapillaris with a temporal resolution sufficient to capture heartbeat-induced blood flow variations~\cite{Puyo2018}. Additionally, the asymmetry of the Doppler spectrum enabled identifying the local axial direction of blood flow.


We used a scanning swept-source OCT instrument built by Heidelberg Engineering with the same optical system as the commercialized Spectralis, except for the light source, scanners, and anti-reflection coatings. A customized $\SI{1}{\mega\hertz}$ microelectromechanical system vertical cavity surface emitting laser (MEMS-VCSEL, Thorlabs) was used as light source, with a central wavelength of $\SI{1060}{\nano\meter}$ and bandwidth of $\SI{90}{\nano\meter}$, yielding an axial resolution of $\SI{5.5}{\micro\meter}$ in air. The galvanometer scanners (dynAXIS XS, SCANLAB GmbH) performed full circles B-scans with a diameter of 6 degrees at a fixed location and were constituted of 500 A-scans.
B-scans were recorded continuously for $\SI{4}{\second}$ at a B-scan rate of $\SI{2}{\kilo\hertz}$, allowing the sampling of frequency shifts up to $f_{\rm Nyquist} = \SI{1}{\kilo\hertz}$. The size of the beam on the pupil was $\SI{1.7}{\milli\meter}$, corresponding to a diffraction limited lateral resolution of $\SI{5}{\micro\meter}$.
The macular region of two subjects was imaged. Experimental procedures adhered to the tenets of the Declaration of Helsinki and written informed consent was obtained from all subjects. All experiments were performed in accordance with relevant guidelines and regulations. Compliance with all relevant safety rules was confirmed by the responsible safety officer.

The data was pre-processed on the acquisition card (AQOCT Solution, Acqiris SA) through the following main steps: finite impulse response filtering, digital k-space resampling, background subtraction, dispersion compensation, and fast Fourier transformation.
B-scans are then segmented to align the RPE at a fixed depth, and the phase instability caused by bulk motion is compensated by subtracted from all B-scans the cumulative phase difference between consecutive B-scans.
The subsequent treatment consists in processing the $(x, z, t)$ OCT data like $(x, y, t)$ holograms are processed in LDH, and is therefore termed laser Doppler OCT.
A time-frequency analysis is performed with a sliding short-time window of 128 consecutive B-scans, corresponding to a temporal resolution of $\SI{64}{\milli\second}$. 
The short-time window of B-scans is Fourier transformed along the temporal dimension, and the power spectrum density is computed from the squared magnitude of the Fourier transform~\cite{Puyo2018}. Blood flow-contrasted power Doppler images are computed by integrating the power spectrum density over a selected frequency range, typically 0.1 to 1 kHz. It should be noted that both speckle decorrelation and Doppler-shifted light contribute to the signal conventionally referred to as power Doppler. Images were generally of slightly higher quality when using the OCT amplitude only. The complex data must however be used to determine for the directional images demonstrated later.
Finally, to avoid power Doppler variations induced by changes in OCT brightness, the local power Doppler was normalized pixel-wise with the OCT amplitude.


\begin{figure}[t!]
\centering
\includegraphics[width = 1\linewidth]{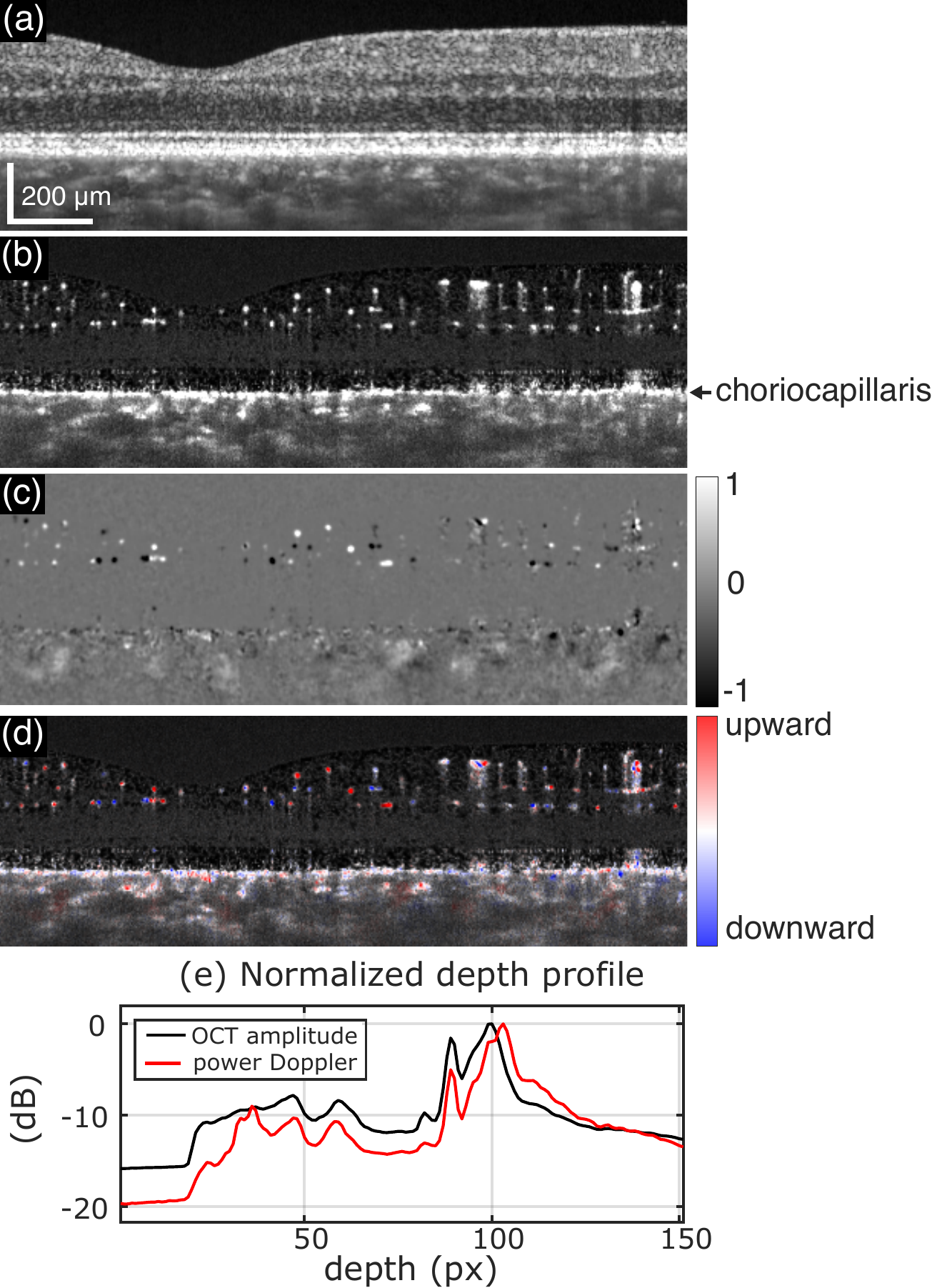}
\caption{Blood flow contrast from the analysis of 128 consecutive B-scans recorded at 2 kHz B-scan rate.
(a) Average reflectivity (log scale).
(b) Power Doppler integrated between 0.1 and 1 kHz.
(c) Power Doppler asymmetry between 0.1 and 1 kHz.
(d) Directional blood flow obtained by combining the previous two images.
(e) Depth profile of the OCT amplitude and power Doppler.
Dynamic movie for the $\SI{4}{\second}$ measurement in \textcolor{blue}{\href{https://youtu.be/iTU-GYysvAQ}{Visualization 1}}.
}
\label{1_Images}
\end{figure}

The outcome of the laser Doppler OCT processing is demonstrated in Fig.~\ref{1_Images} for a short-time window of 128 consecutive circular B-scans.
The average reflectivity is shown in logarithmic scale in Fig.~\ref{1_Images}(a), and the blood flow image resulting from the power Doppler integration over 0.1 to 1 kHz is shown in Fig.~\ref{1_Images}(b). Retinal and choroidal vessels appear with a bright contrast on the power Doppler image. The choriocapillaris layer especially gives a strong blood flow signal.
In Fig.~\ref{1_Images}(c), the power Doppler asymmetry, i.e., the difference between the positive and negative parts of the spectrum broadening, reveals the local direction of blood flow. Many retinal vessels show a clear axial component, showing the sensitivity of the method to slight angles. No axial flow direction is however detected in almost all choroidal vessel.
The images of the power Doppler sum and difference are combined in one color Doppler image in Fig.~\ref{1_Images}(d).
We used the processing developed to generate a directional contrast for LDH~\cite{Puyo2021Directional}, based on a Hue-Saturation-Value combination. In short, the red or blue Hue is given by the respectively positive or negative sign of the power Doppler difference. The Saturation is given by the absolute value of the power Doppler difference, and the Value is given by the power Doppler. The contrast of both the power Doppler sum and difference images is automatically adjusted by saturating the pixels with the lowest and brightest intensities beyond fixed arbitrary thresholds.
The resulting color Doppler or directional image is able to show at the same time the local blood flow and the detected axial direction of the flow.
Finally, the averaged logarithmic profiles of the OCT amplitude and of the power Doppler are shown in Fig.~\ref{1_Images}(d). The two curves follow one another relatively closely, but the OCT amplitude reaches its maximum at the RPE while the power Doppler peaks at the choriocapillaris, as was already observed with OCT-A\cite{Chu2021}.
The dynamic movie for the $\SI{4}{\second}$ measurement is shown in \textcolor{blue}{\href{https://youtu.be/iTU-GYysvAQ}{Visualization 1}}.
Different features can be seen over time, due to the eye drifting and the absence of active tracking.

\begin{figure}[t!]
\centering
\includegraphics[width = 1\linewidth]{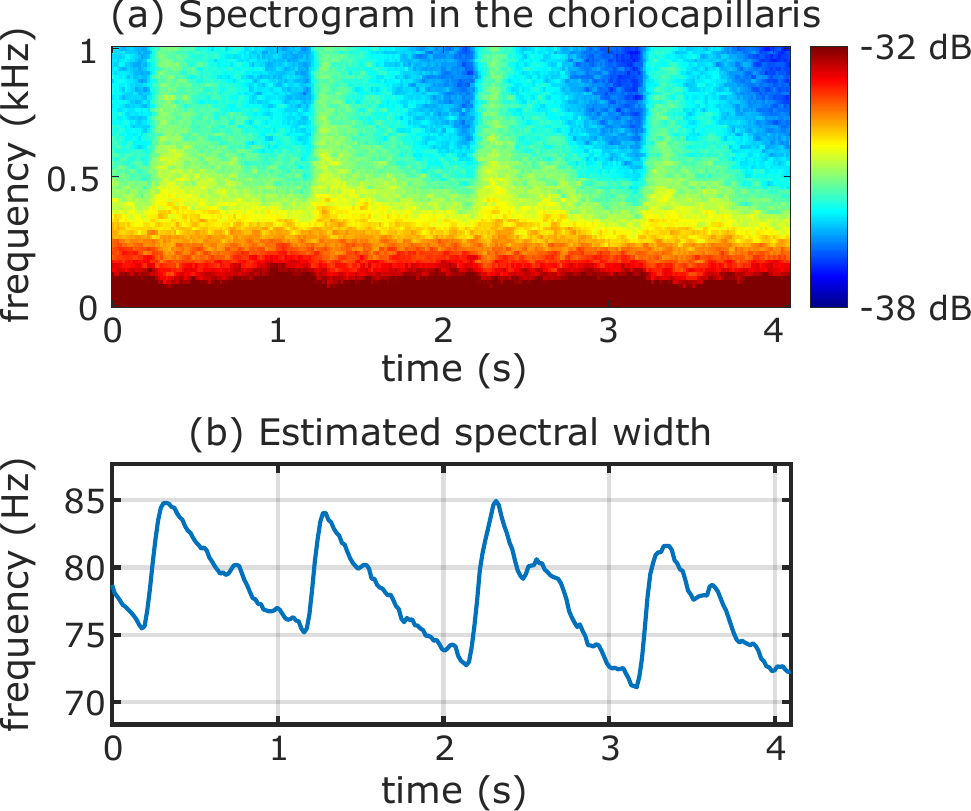}
\caption{Measurement of heartbeat-induced blood flow variations. (a) Spectrogram of light scattered in the choriocapillaris revealing pulsatile changes of spectrum broadening. (b) The dynamic standard deviation of the power spectrum density yields a characterization of choriocapillaris blood velocity in physical units (Hz).
Dynamic movie shown in \textcolor{blue}{\href{https://youtu.be/haffN9vaVoc}{Visualization 2}}.
}
\label{2_Spectrogram_Plot}
\end{figure}

Pulsatile variations of blood flow are investigated in Fig.~\ref{2_Spectrogram_Plot}. A measurement without any micro-saccade was selected. The sliding window analysis was again performed with windows constituted of 128 consecutive B-scans. The choriocapillaris layer was detected from the peak of power Doppler. The power spectrum density in the choriocapillaris layers was spatially averaged over the B-scan width and its temporal variations are presented as a spectrogram in Fig.~\ref{2_Spectrogram_Plot}(a). Heartbeat-induced flow variations are revealed in the spectral fluctuations.
The standard deviation of the power spectrum density is computed for each short-time window to estimate the dynamic changes in spectrum broadening~\cite{Zhao2000}, and the resulting plot is shown in Fig.~\ref{2_Spectrogram_Plot}(b). The dynamic movie is shown in \textcolor{blue}{\href{https://youtu.be/haffN9vaVoc}{Visualization 2}}.
Pulsatile variations of blood flow in the choriocapillaris meshwork are revealed and quantified in Hz. A stable and reproducible waveform with a small amplitude of variations is obtained.

\begin{figure}[t!]
\centering
\includegraphics[width = 0.97\linewidth]{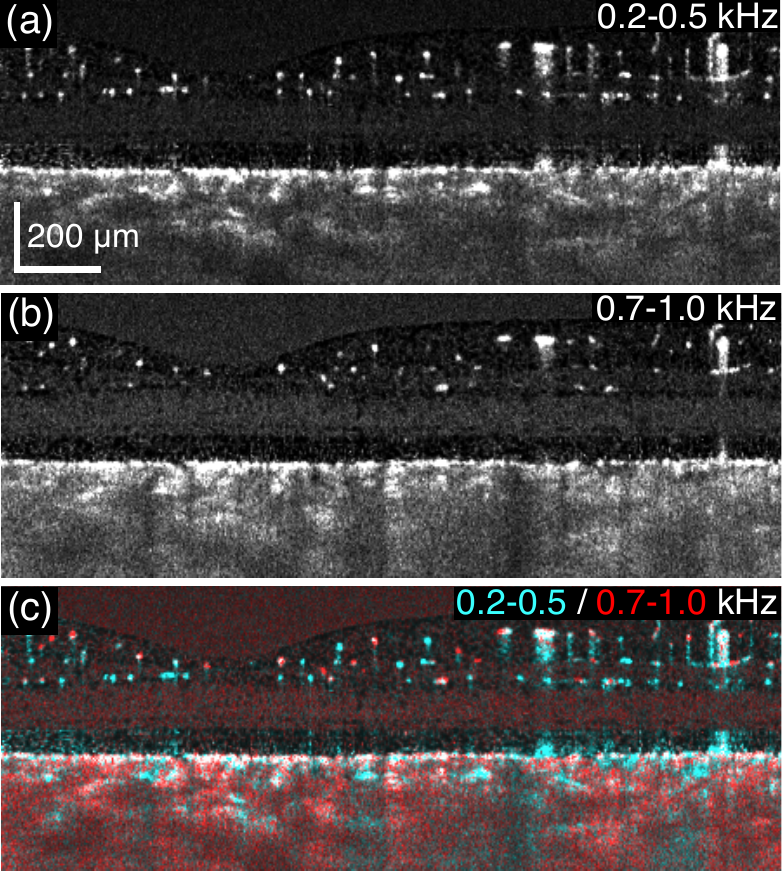}
\caption{Power Doppler images at different frequency bands. (a) 0.2-0.5 kHz. (b) 0.7-1 kHz. (c) Composite image obtained by merging low and high frequency power Doppler images in cyan and red, respectively. Vessels appear in cyan or red depending on their flow velocity and geometry.
Power Doppler frequency movie in \textcolor{blue}{\href{https://youtu.be/TAM-rmArgkc}{Visualization 3}}.
Dynamic composite movie in \textcolor{blue}{\href{https://youtu.be/SB49PvZw16E}{Visualization 4}}.
}
\label{3_FrequencyBands}
\end{figure}

One useful feature of power Doppler is the possibility to form images for different frequency bands~\cite{Puyo2019a}. For a short-time window of 128 B-scans, power Doppler images computed for the 0.2-0.5 kHz and 0.7-1 kHz frequency ranges are shown in Fig.~\ref{3_FrequencyBands}(a-b). The power Doppler images obtained with a sliding frequency window are shown in \textcolor{blue}{\href{https://youtu.be/TAM-rmArgkc}{Visualization 3}}. The power Doppler images from the low and high frequency bands were fused into a color composite image in cyan and red, respectively, shown in Fig.~\ref{3_FrequencyBands}(d).
Most retinal vessels appear exclusively in either the low- or high-frequency power Doppler images and are thus visible in cyan or red in the combined image. However, some larger retinal vessels are visible in both the low- and high-frequency power Doppler images, appearing white in the composite image. We assume that this is due to aliasing, which shifts energy from higher frequency bands to lower frequency bands.
Higher velocities cause faster decorrelation or higher Doppler shifts, which generate power Doppler signals at higher frequencies. This provides a qualitative assessment of velocities in vessels. However, the frequency shift is also influenced by the orientation of the vessels relative to the optical axis, as a higher Doppler shift also shifts the power Doppler to higher frequencies.

In the retina, blood flow can be measured by summing the flow from all branches of the central retinal artery, as the supply comes from a single vessel. However, applying this approach to the choroid is far more challenging. Choroidal blood flow is supplied by multiple posterior ciliary arteries, which vary in number and location from one eye to another. As a result, measuring choroidal blood flow at the capillary level, where it supplies oxygen and nutrients, is likely a more relevant approach. Our method estimates in physical units the spectral broadening of light scattered within the choriocapillaris layer.
Since faster flow in the choriocapillaris results in a larger spectral broadening, it can be reasonably expected that this index would increase with a better vascularization of the choriocapillaris.
We aimed to obtain laser Doppler OCT measurements representative of the choriocapillaris ensemble. By spatially averaging over enough A-scans, the effects of varying vessel geometries are minimized, and the obtained flow value could be representative of the overall meshwork. As can be observed in \textcolor{blue}{\href{https://youtu.be/haffN9vaVoc}{Visualization 2}}, eye movements result in different parts of the choriocapillaris being imaged over time, but steady pulsatile flow profiles could nonetheless be obtained. This observation supports the idea that the choriocapillaris may be sufficiently uniform, so that the values obtained are similar across different areas, and could therefore reflect the entire meshwork.
Although we did not observe aliasing, the B-scan sampling rate could be increased if necessary by reducing the number of A-scans per B-scan. This would however in turn limit the extension of the choriocapillaris area being imaged, which could be mitigated by actively displacing the B-scan during the measurement.
This approach has the potential to provide valuable insights into choriocapillaris perfusion, offering complementary information to structural metrics like vascular density and flow voids. This could be particularly useful for understanding decreased blood flow, progressive degeneration of the choriocapillaris, and choroidal neovascularization.
In comparison to LDH, axial sectioning is critical in the success of laser Doppler OCT to study the choriocapillaris. The mixing of signals from retinal and choroidal depths indeed prevents LDH from revealing capillary. Moreover, laser Doppler OCT benefits from the reduction of multiple scattering and better interpretability of the signal inherent to OCT.
Finally, laser Doppler OCT could be advantageously implemented on clinical high-speed Fourier-domain point-scanning OCT systems. As measurements are provided in physical units, this technique could, in principle, ultimately allow for inter-individual and inter-instrument comparisons.

This work was carried out without active tracking to maintain the B-scan at a fixed retinal location. Since compensating for eye movements through registration is not possible, the sliding short-time window must be short enough to ensure that signal variations are due to blood flow rather than imaging different retinal features. Micro-saccades were found to be disrupting the signal because they cause much stronger changes than blood flow, but they can be detected. Slower eye movements seem however not to be an issue. This should depend on the speckle size and investigated range of velocities. Smaller blood vessels are more challenging to detect because they require a smaller speckle size, making them more susceptible to eye displacement, and longer short-time windows, as blood flow is slower. Active eye-tracking could help capture slower velocities and smaller vessels. It would also enable following blood flow changes in the same retinal vessels over cardiac cycles, and therefore measuring pulsatile flow in retinal capillaries. In larger vessels, it would also allow performing an arteriovenous differentiation based on flow variations during cardiac cycles~\cite{Puyo2019b}.


In 2019, Migacz et al. showed with an experimental OCT device using a 1.6 MHz Fourier domain mode-locked (FDML) laser that the choriocapillaris is better revealed with faster temporal sampling and found that $\SI{2}{\kilo\hertz}$ is an appropriate B-scan rate~\cite{Migacz2019}. Our findings are consistent with this observation, as we observe only little signal above $\SI{1}{\kilo\hertz}$ in the spectrum of light scattered in the choriocapillaris. This stands in contrast to the rate of a few hundred Hz that is typically used in OCTA~\cite{Wei2019}.
Laser Doppler OCT will however largely benefit from higher A-scan rates. This would allow scanning larger fields of view to average more of the choriocapillaris. More importantly, for a given field of view, this would allow sampling larger velocities, i.e., measure blood flow in larger retinal and choroidal vessels. Faster B-scan sampling rates would also allow identifying the axial direction of blood flow in the large choroidal vessels, which would allow a better understanding of the flow circulation in this 3D vascular compartment.
Moreover, imaging faster velocities would also allow performing the same differentiation of choroidal arteries and veins based on velocity as in LDH. It was indeed found with LDH that blood flow in choroidal veins is slower and typically revealed over the frequency range of 1-6 kHz, whereas choroidal arteries have faster flows, leading to a broadening of the spectrum up to 30 kHz or more~\cite{Puyo2019a}. In principle, the same differentiation is possible with OCT, although the exact frequency ranges may differ since scattering properties of light detected in OCT and LDH are different.

Capillaries showing a power Doppler signal without directional contrast are revealed through speckle decorrelation. The presence of directional contrast in some retinal capillaries reveals on the other hand the contribution of Doppler shifts. Therefore, both speckle decorrelation and Doppler shifts contribute to the detected spectrum. A key advantage of the power Doppler approach is its ability to account for both phenomena, providing more insight in eye dynamic light scattering.
Retinal vascular map supplemented with a color Doppler contrast could be used to improve retinal blood flow models, which is a crucial step toward better understanding ocular blood flow and advancing personalized healthcare~\cite{Sun2023}. While current methods generate detailed maps of the retinal capillary network, the flow circulation between different vascular plexuses is a missing information that could be obtained with the directional contrast. By scanning volumes and producing en-face images, laser Doppler OCT could capture this critical directional information. This was not performed in the current study for lack of an active eye tracking system.



To summarize, we used a high-speed Fourier-domain point-scanning OCT device to capture B-scans continuously at the same location. The data were processed using a sliding short-time window Fourier analysis, and the standard deviation of the power spectrum density was used to measure in physical units the spectral broadening of light scattered from the choriocapillaris. This index could be expected to reflect the perfusion of the overall choriocapillaris meshwork. Additionally, dynamic changes could be evaluated with sufficient temporal resolution to capture pulsatile changes. Finally, the Doppler spectral asymmetry enabled us to determine the axial direction of blood flow. As this technique will benefit from higher A-scan rates and can be implemented on conventional high-speed scanning OCT devices, it holds promise for further development.

\section*{Funding}
Deutsche Forschungsgemeinschaft (HU 629/6-1).
Bundesministerium f\"ur Bildung und Forschung (BMBF 13N15432).

\section*{Disclosures}
LK: Heidelberg Engineering GmbH (E).

\bibliographystyle{unsrt}
\bibliography{./Bibliography}

\end{document}